\documentclass[twocolumn]{aastex63}

\usepackage[T1]{fontenc}
\usepackage{ae,aecompl}
\usepackage{graphicx}
\usepackage{amsmath}
\usepackage{amssymb}
\usepackage{supertabular}
\usepackage{color}
\usepackage{longtable}
\usepackage{booktabs}

\shorttitle{Compact remnants in CCSNe and the lower mass gap}
\shortauthors{Liu et al.}

\begin{document}

\title{Final compact remnants in core-collapse supernovae from 20 to 40 $M_\odot$: the lower mass gap}

\author[0000-0001-8678-6291]{Tong Liu}
\affiliation{Department of Astronomy, Xiamen University, Xiamen, Fujian 361005, China\\ tongliu@xmu.edu.cn, lixue@xmu.edu.cn, msun88@xmu.edu.cn}

\author{Yun-Feng Wei}
\affiliation{Department of Astronomy, Xiamen University, Xiamen, Fujian 361005, China\\ tongliu@xmu.edu.cn, lixue@xmu.edu.cn, msun88@xmu.edu.cn}

\author{Li Xue}
\affiliation{Department of Astronomy, Xiamen University, Xiamen, Fujian 361005, China\\ tongliu@xmu.edu.cn, lixue@xmu.edu.cn, msun88@xmu.edu.cn}

\author[0000-0002-0771-2153]{Mou-Yuan Sun}
\affiliation{Department of Astronomy, Xiamen University, Xiamen, Fujian 361005, China\\ tongliu@xmu.edu.cn, lixue@xmu.edu.cn, msun88@xmu.edu.cn}

\begin{abstract}
A mass paucity of compact objects in the range of $\sim 2-5 ~M_\odot$ has been suggested by X-ray binary observations, namely, the ``lower mass gap''. Gravitational wave detections have unlocked another mass measurement method, and aLIGO/Virgo has observed some candidates in the gap. We revisit the numerical simulations on the core-collapse supernovae (CCSNe) for $\sim 20-40~M_\odot$ progenitor stars with different initial explosion energies. As a result, the lower explosion energy naturally causes more efficient fallback accretion for low-metallicity progenitors, and then the newborn black holes (BHs) in the center of the CCSNe can escape from the gap, but neutron stars cannot easily collapse into BHs in the gap; nevertheless, the final remnants of the solar-metallicity progenitors stick to the gap. If we consider that only drastic CCSNe can be observed and that those with lower explosion energies are universal, the lower mass gap can be reasonably built. The width and depth of the gap are mainly determined by the typical CCSN initial explosion energy and metallicity. One can expect that the future multimessenger observations of compact objects delineate the shape of the gap, which might constrain the properties of the CCSNe and their progenitors.
\end{abstract}


\section{Introduction}

Discoveries of stellar-mass black holes (BHs) and neutron stars (NSs) are resulting from the observations of compact X-ray binaries. The masses of more than 20 BHs and more than 70 NSs have been determined in these binaries to date. \citet{Bailyn1998} first noticed that there are almost no compact remnants in the range of $\sim 2-5~M_\odot$, which was later verified \citep[e.g.,][]{Ozel2010,Farr2011} and named the ``lower mass gap'' (or the ``first mass gap''). Three massive NSs, PSR J1614-2230, PSR J0348+0432, and MSP J0740+6620, were detected during the last decade \citep{Demorest2010,Antoniadis2013,Cromartie2020}. Their masses can be measured by the ``Shapiro delay'' effects caused by their white dwarf companions. According to measurements with the magnesium triplet lines, \citet{Linares2018} reported an NS, PSR J2215+5135, with approximately $2.3~M_\odot$, accompanied by a low-mass, nondegenerate star. However, the values of these NS masses exceed the lower limit of the gap and challenge the equation of state at supranuclear densities \citep[e.g.,][and references therein]{Li2020}. Recently, \citet{Thompson2019} discovered a $\sim 3~M_\odot$ BH candidate in a noninteracting low-mass binary system. In the aLIGO/Virgo detections, the compact remnants in two NS-NS merger events, GW170817 \citep{Abbott2017} and GW190425 \citep{Abbott2020a}, and one of the objects in a binary merger event, GW190814 \citep{Abbott2020b}, are definitely in the gap. Thus, the questions we must answer are whether the gap exists and how large the gap is in the mass distribution of the compact objects.

Two possible approaches, i.e., mergers of compact binary or multibody systems \citep[see, e.g.,][]{Drozda2020,Fragione2020,Yang2020,Zevin2020} and fallback hyperaccretion in collapsars \citep[see, e.g.,][]{Fryer2001,Heger2002,Heger2003,Zhang2008,Belczynski2012,Fryer2012,Liu2018,Chan2020}, can significantly increase the values of the compact remnant masses.

\citet{Colgate1971} first proposed the supernova (SN) fallback mechanism. In this scenario, regardless of whether or not the SN explosions are sufficiently energetic to disperse the gas of progenitor stars, some materials may still fallback onto the cores, increasing their masses, changing their rotations and magnetic fields, and potentially transferring an NS to a BH. Numerous studies in the literature have calculated or simulated the complicated dynamics and effects of the fallback processes in core-collapse SNe \citep[CCSNe, e.g.,][and references therein]{Bisnovatyi-Kogan1984,Woosley1989,Chevalier1989,Woosley1995,Fryer1999,Fryer2009,MacFadyen2001,Zhang2008,Moriya2010,Dexter2013,Wong2014,Perna2014,Branch2017}. The intensity of the fallback depends on the initial explosion energy, then the material apportionment is further achieved between the compact object accretion and SN ejecta.

For the lower mass gap, \citet{Belczynski2012} simulated that there were almost no remnants in the gap during the rapid explosions of CCSNe conforming to the previous X-ray binary observations, while delayed explosions resulted in a continuous compact object mass distribution. If the gap exists, most of compact objects with an initial mass of $\sim 2-5~M_\odot$ should jump out of the gap. Furthermore, most NSs cannot collapse into BHs and thus growth ceases below the lower limit of the gap. In this work, we revisit the physical processes of the fallback accretion and core-collapse and discuss the possible origin of the lower mass gap in the multimessenger era.

In this paper, the initial explosion energy effects on the fallback accretion processes in the CCSN scenario are simulated by the Athena++ code. Then, we describe the variant shapes of the gap regions within the different distributions of the initial explosion energy. In Section 2, we present the simulation method and main results on the relation between the masses of the final compact remnants and the masses and metallicities of the progenitors. The shapes of the mass distributions of the compact objects with the different initial explosion energy distributions are shown in Section 3. Conclusions are summarized in Section 4 with some discussion.

\begin{figure*}
\includegraphics[width=0.5\linewidth]{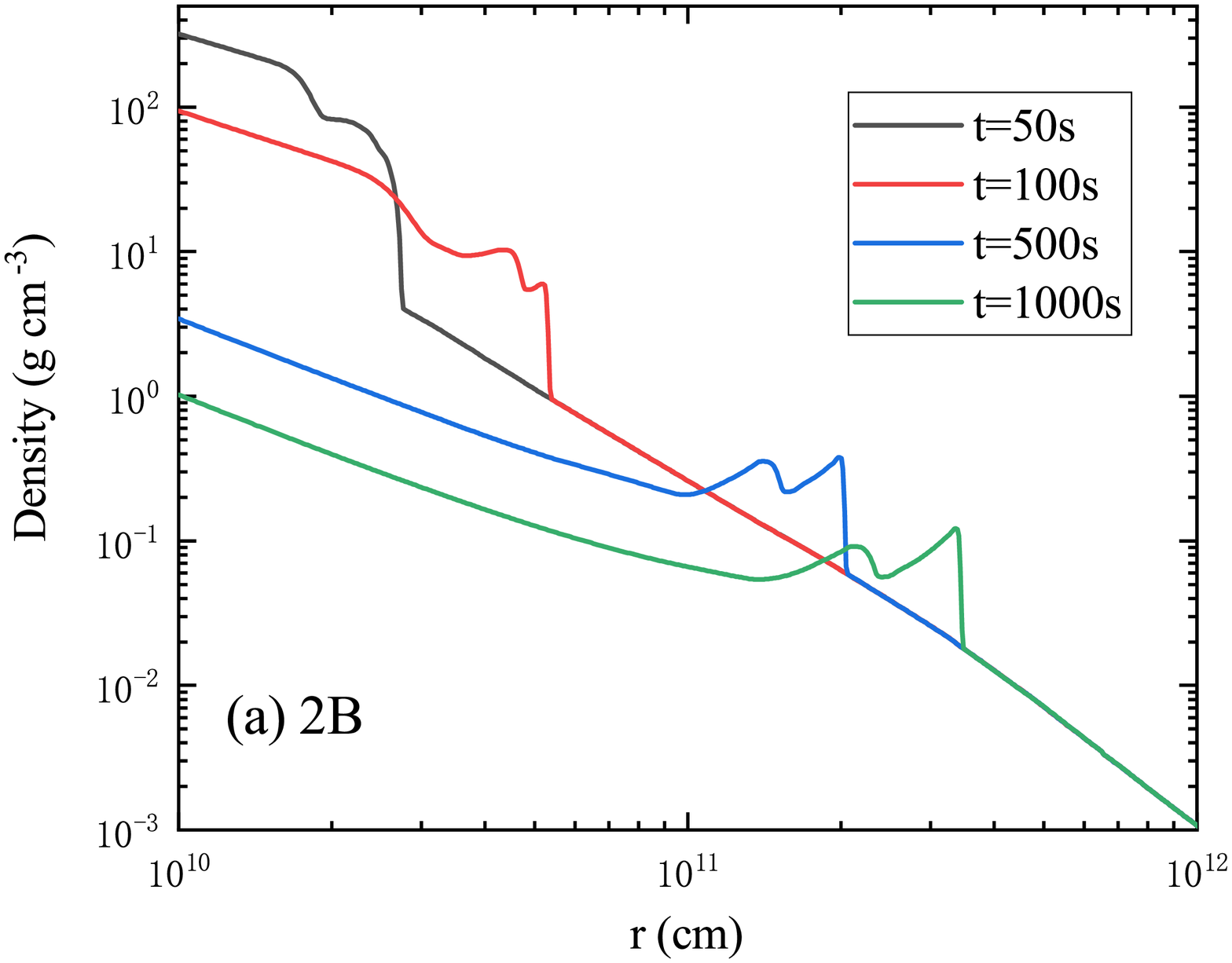}
\includegraphics[width=0.5\linewidth]{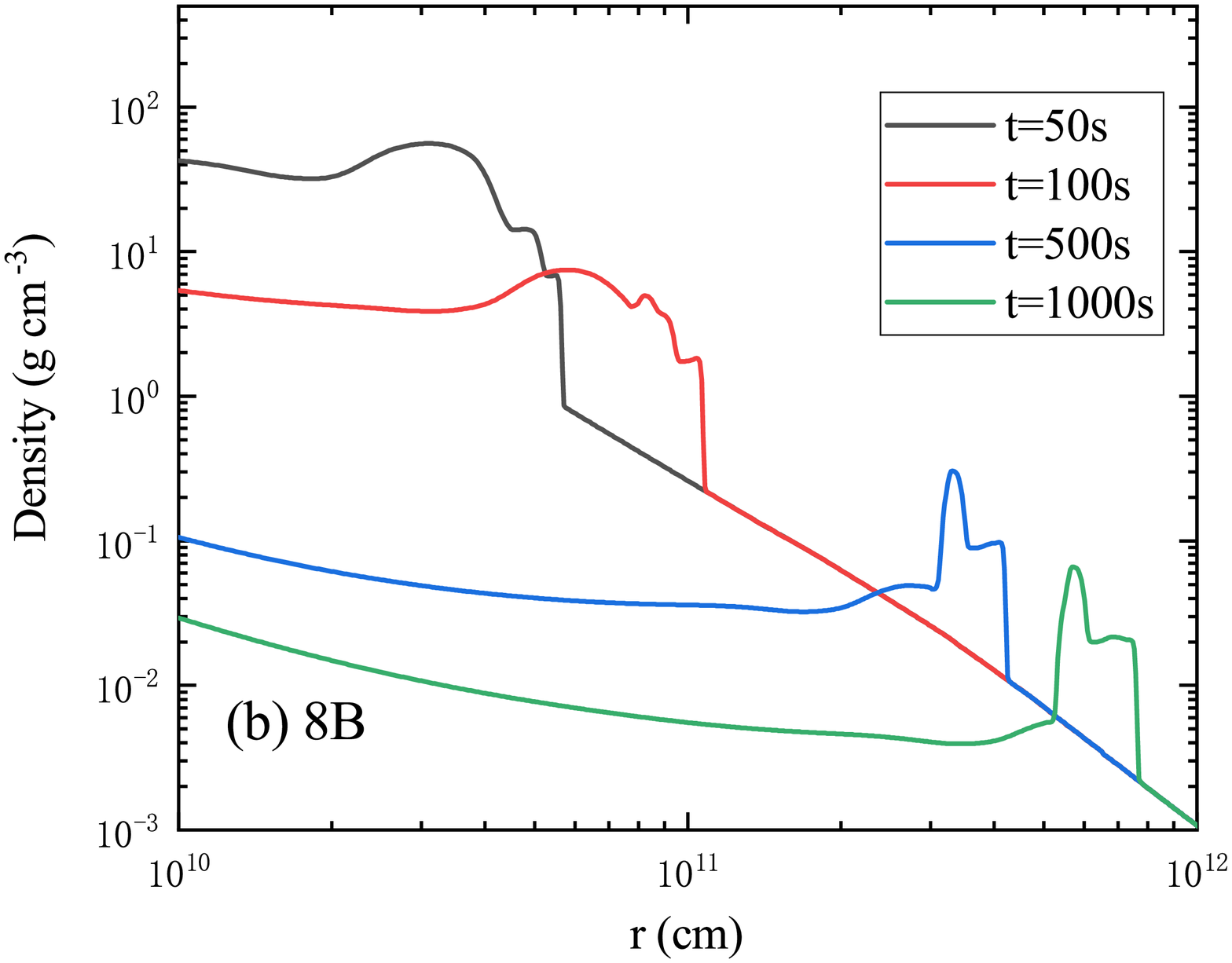}
\includegraphics[width=0.5\linewidth]{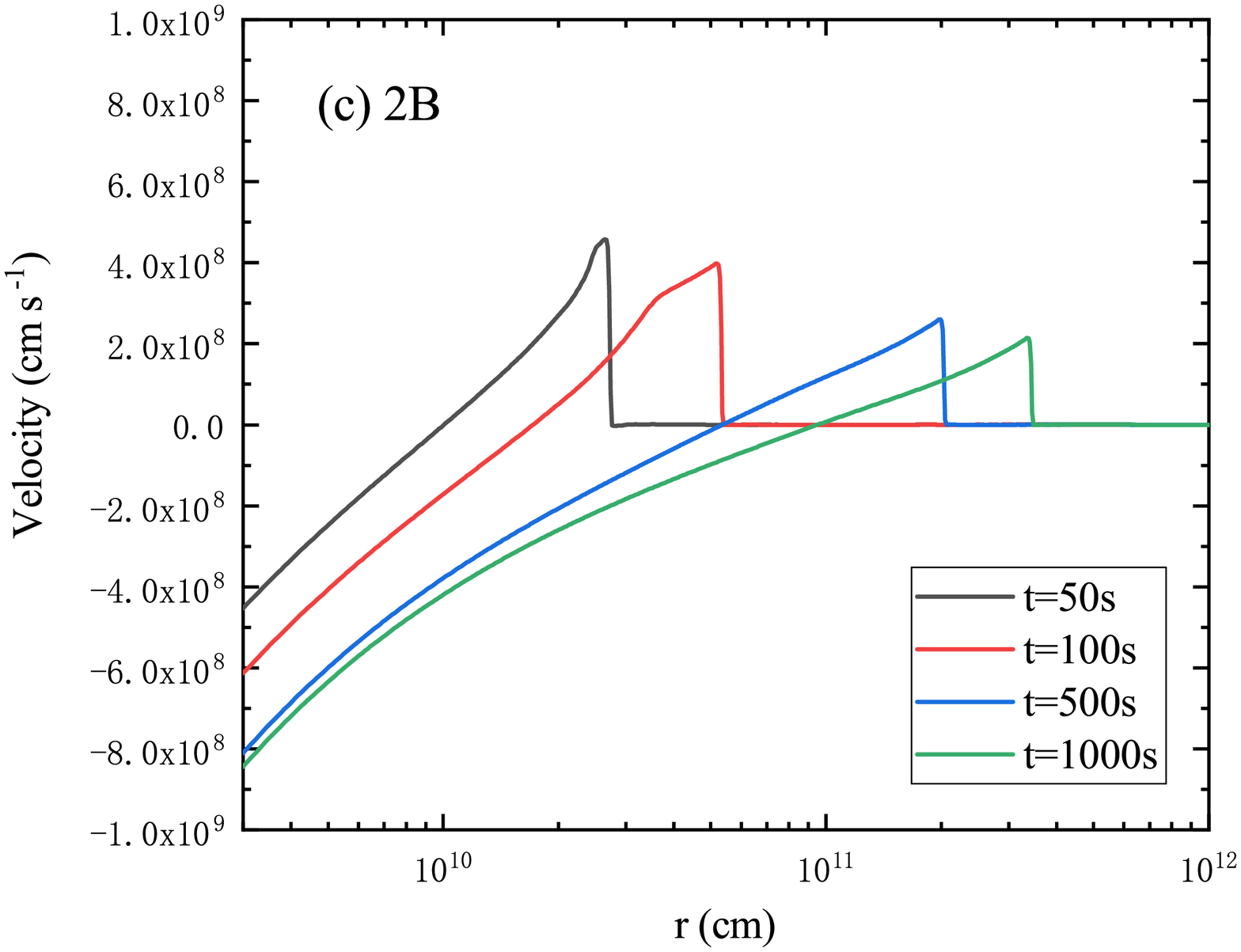}
\includegraphics[width=0.5\linewidth]{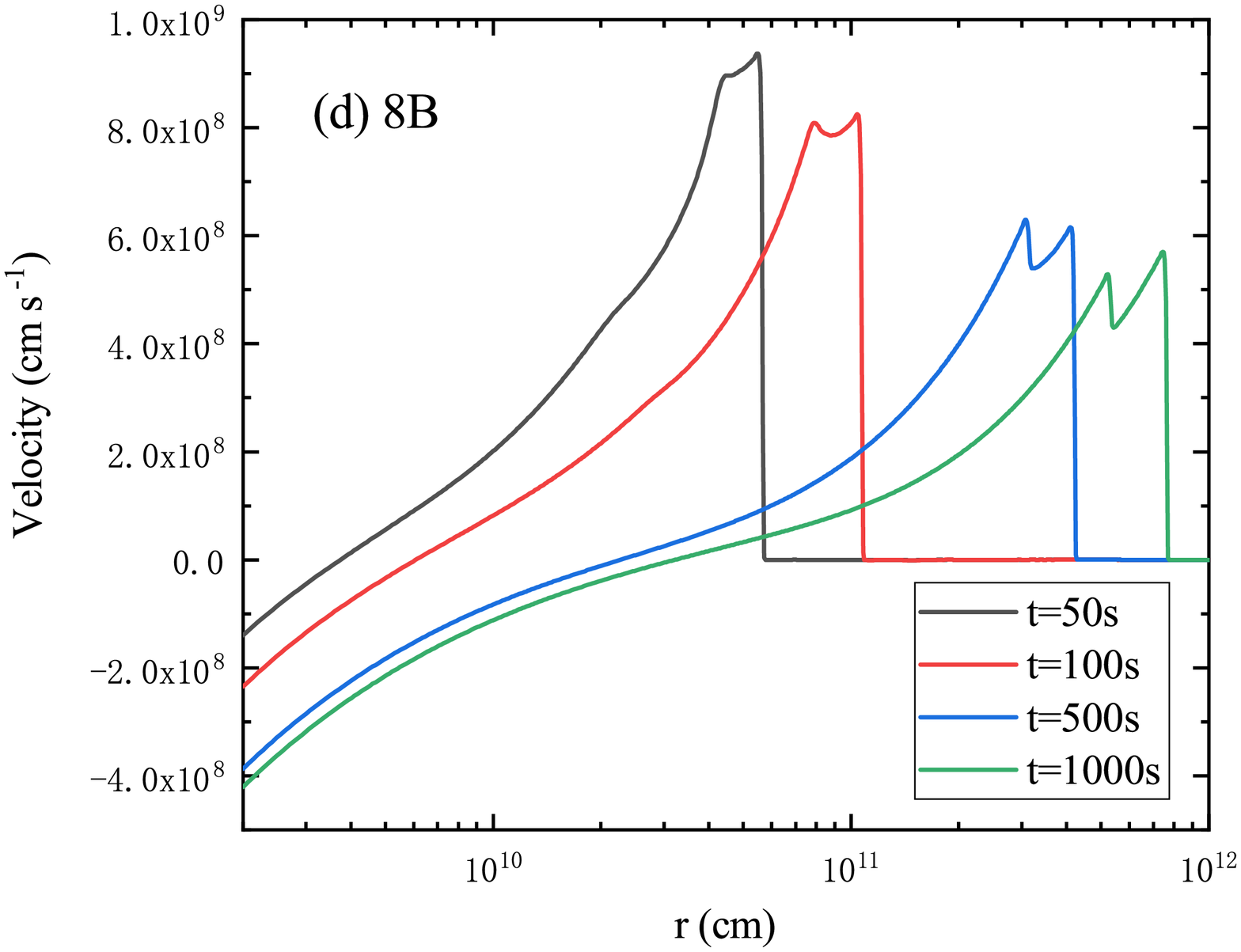}
\includegraphics[width=0.5\linewidth]{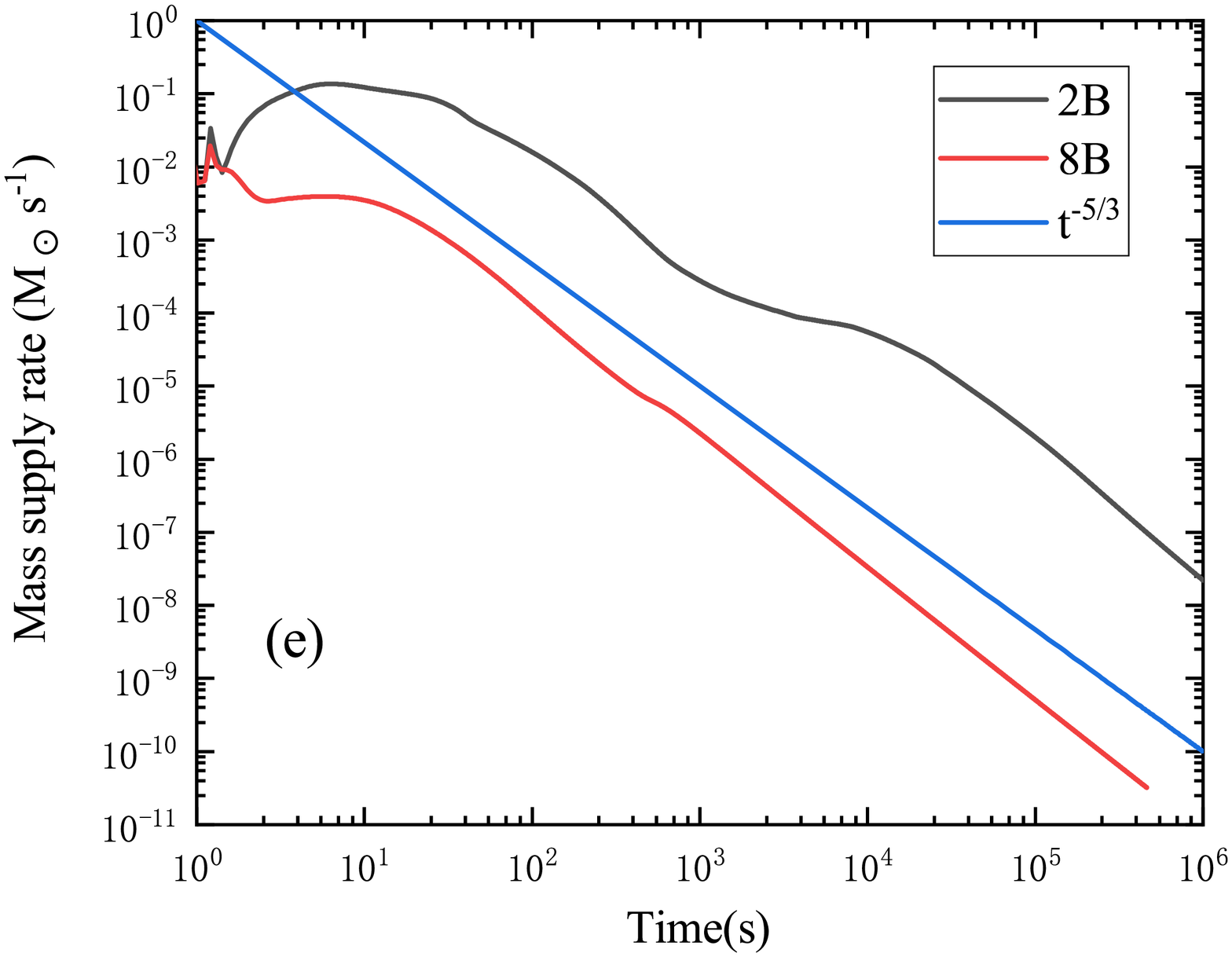}
\includegraphics[width=0.5\linewidth]{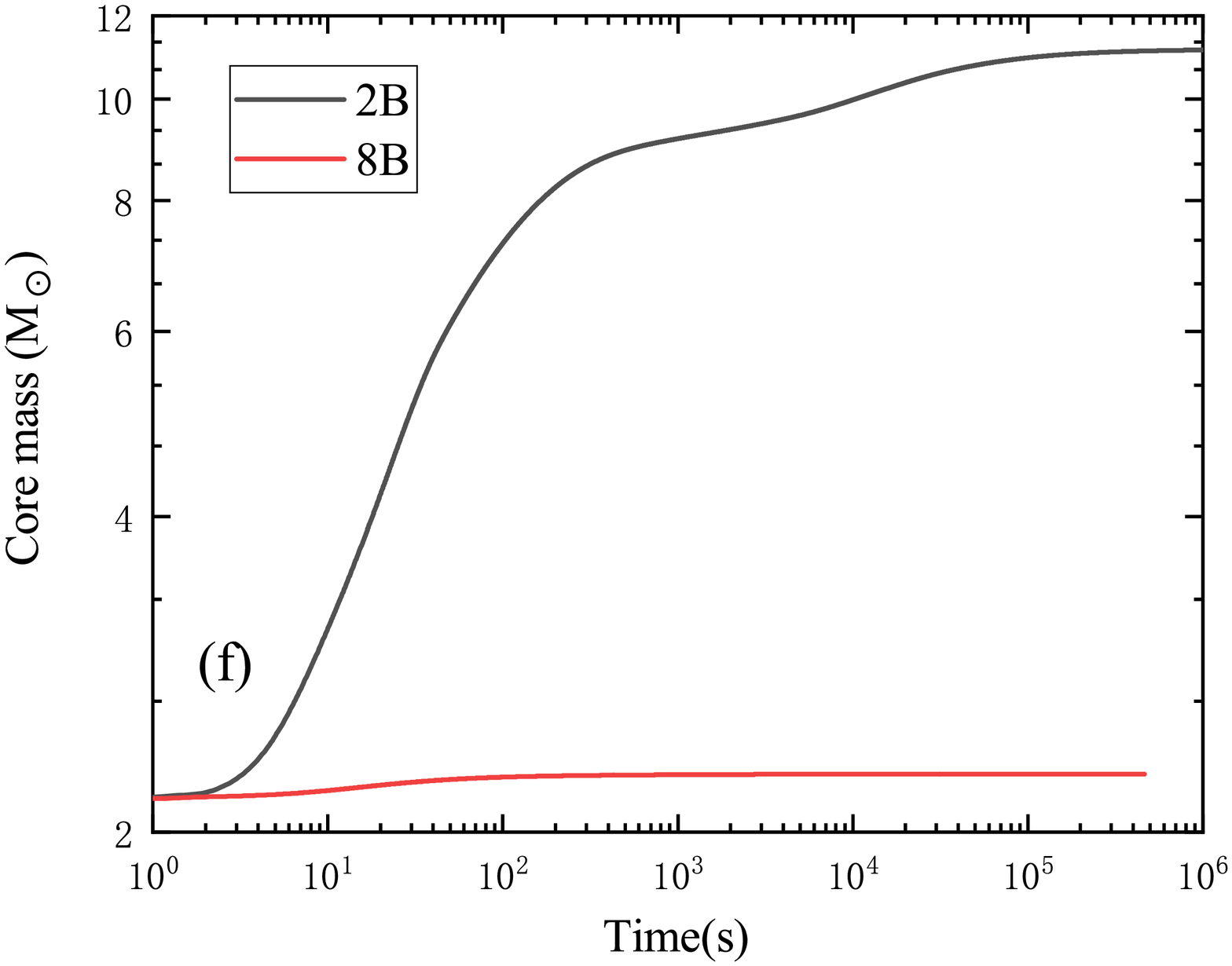}
\caption{Profiles of density and velocity at 50, 100, 500, and 1000 s and time evolutions of mass supply rate and core mass for $E$ = 2B and 8B with progenitor mass $30~M_\odot$ and $Z/Z_\odot$ = 0.01.}
\label{fig:profile-evolution}
\end{figure*}

\section{BH mass growths in CCSNe}

\subsection{Method}

We mainly refer to the method of \cite{Woosley1995} to carry out spherically symmetric explosion simulations. We consider here the simulated explosions of the massive progenitor stars ($20-40~M_{\odot}$) with metallicity values of $Z/Z_\odot$ = 0, 0.01, 0.1, and 1 (where $Z_\odot$ represents the metallicity of the Sun) and with different initial explosion energies ($E$ = 2B, 4B, 6B, and 8B, where 1B = $10^{51}\rm{erg}$). The dynamical evolutions of the explosions in these cases are performed in a series of 1D simulations with the Athena++ code \citep{White2016}. The effects of gravity are included as a user-defined source term in Athena++. A point mass, which is updated by tracking the mass flux across the inner boundary, is placed at the center of the grid to mimic the compact remnant. The gravitational force at a grid cell is calculated with the enclosed mass below the radius of this cell, including the central point mass and gas mass in cells.

Due to the complexity of the CCSN explosions, we follow \cite{Woosley1995} to simplify the initial explosion with a hypothetical moving piston. The piston is located on the surface of the iron core in the center of the progenitor star. With this kind of assumption, the core masses in all most of cases are well below the upper mass limit of observed massive NSs. Thus, there will always be a hard surface to produce the reflecting shock wave for the core-collapse in those cases. Moreover, the initial core masses for the more massive progenitor stars ($> 40~M_\odot$) directly collapse to BHs and are generally larger than approximately $5~M_\odot$. In addition, as in the results of \citet{Zhang2008}, the $<20~M_\odot$ progenitors cannot produce a final remnant with a mass of $>2~M_\odot$ in our simulations, even with the low explosion energy of $\sim$ 1B, so they are both excluded from our solutions.

When the core collapses, the piston moves rapidly inward for a brief period, $\sim 0.45~\rm{s}$, and then abruptly moves outward at a certain small radius with a high velocity and decelerates smoothly until stopping at a large radius, $\sim 10^9~ \rm{cm}$. Such a hypothetical simplified process has been implemented and proven valid in a series of previous works \citep[e.g.,][]{Woosley1995,Woosley2002}. Here, we also follow this approach in all considered cases and ultimately determine the initial explosion condition at the inner boundary.

The structure profiles of progenitor stars used as the initial condition in simulations are all obtained from \citet{Woosely2012}. The simulation for each case is divided into two steps to reflect the initial collapse and the subsequent explosion accompanied by the fallback process. In the first step, the computational domain has an inner boundary at $r=10^9 ~\rm{cm}$, which is the largest radius reached by the hypothetical piston, and an outer boundary is set at the surface of the progenitor star, which is different for each case (from approximately $10^{12}$ cm to $10^{14}$ cm). In this step, the grid has $10^4$ cells with a logarithmic uniform interval for the $r$-direction to diminish the remapping error in the next step. The simulation is run to $0.45~ \rm{s}$ with a unidirectional outflowing inner boundary condition\footnote{It is achieved by inserting a function into the integrator loop of Athena++ to check the direction of the mass flux on the boundary interface and artificially set the flux to zero if the spurious mass inflow from the ghost cell is detected.} to mimic the suction effect resulting from the hypothetical piston moving inward. This reflects a brief free collapse of the star before the explosion begins. After $0.45~\rm{s}$, the piston turns outward, which corresponds to the outward propagation of the blast. According to CCSN simulations \citep[e.g.][]{Burrows2019}, it takes on average approximately 1 s for a typical blast wave to reach $r=10^9 ~\rm{cm}$, which is also the period in which the hypothetical piston moves outward. During this brief period, the gas flow does not change much as the collapse ceases. Therefore, we directly map the results of the first step, which is a snapshot of the star at 0.45 s, to the new grid for the second step.

\begin{figure*}
\includegraphics[width=0.5\linewidth]{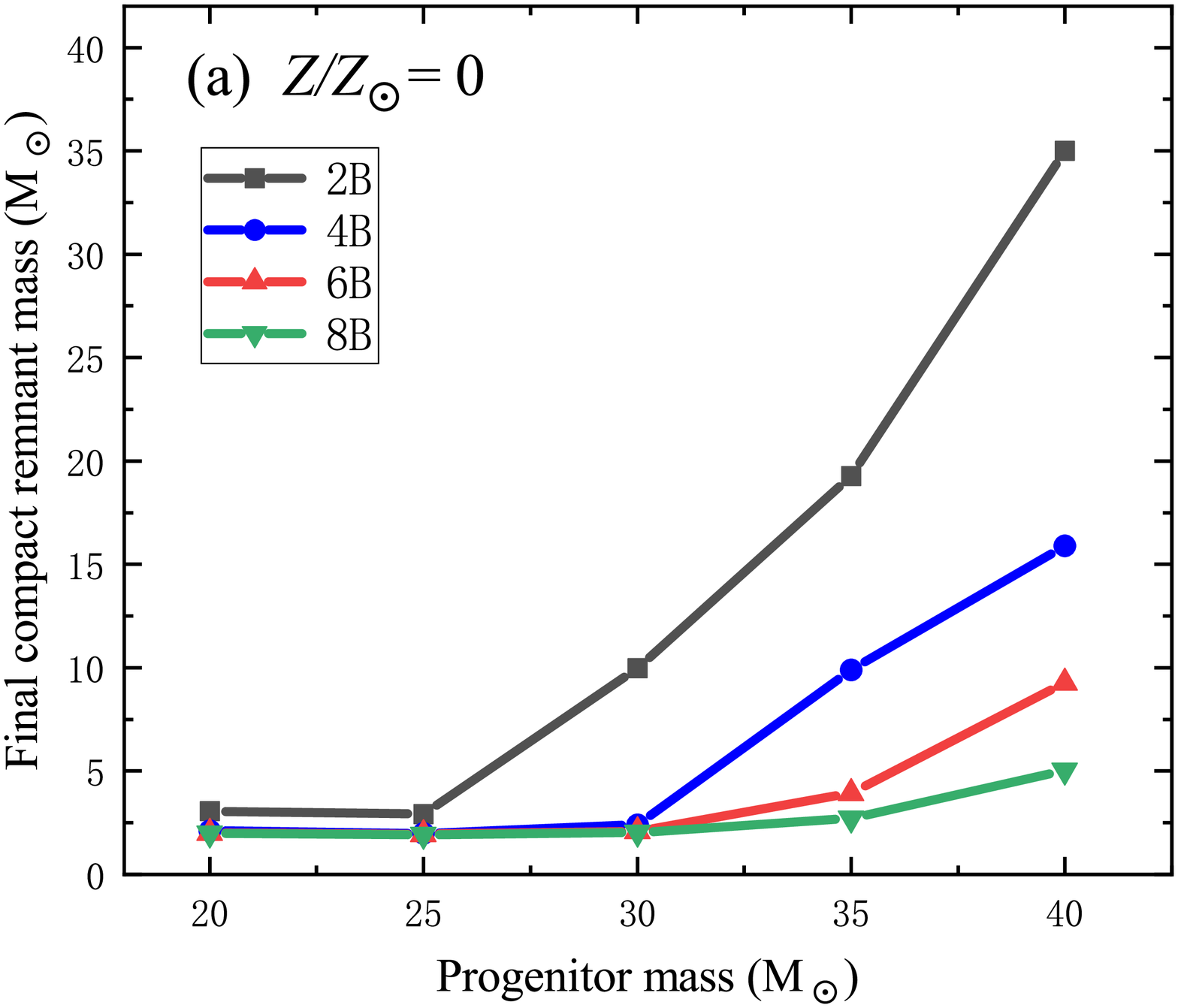}
\includegraphics[width=0.5\linewidth]{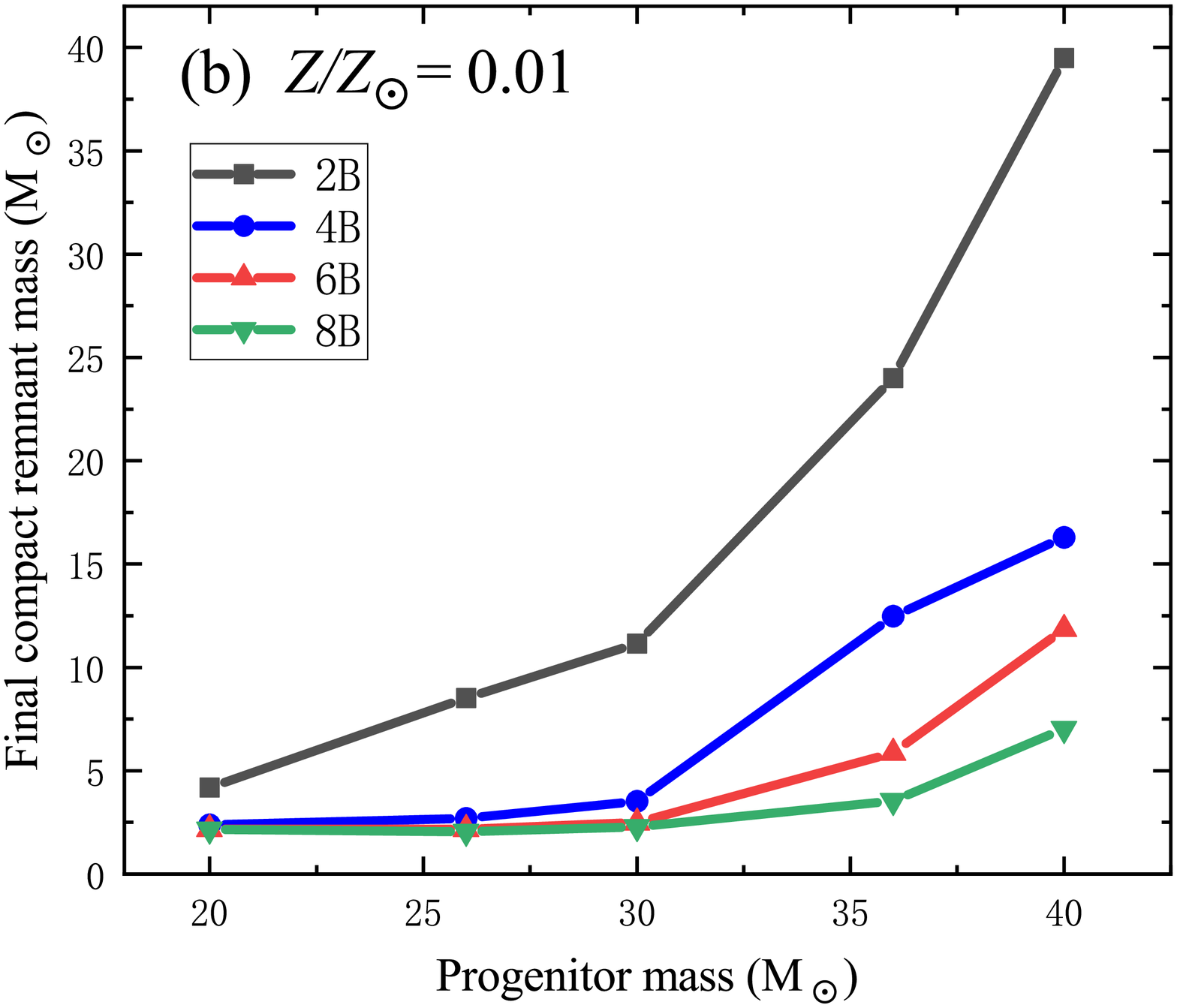}
\includegraphics[width=0.5\linewidth]{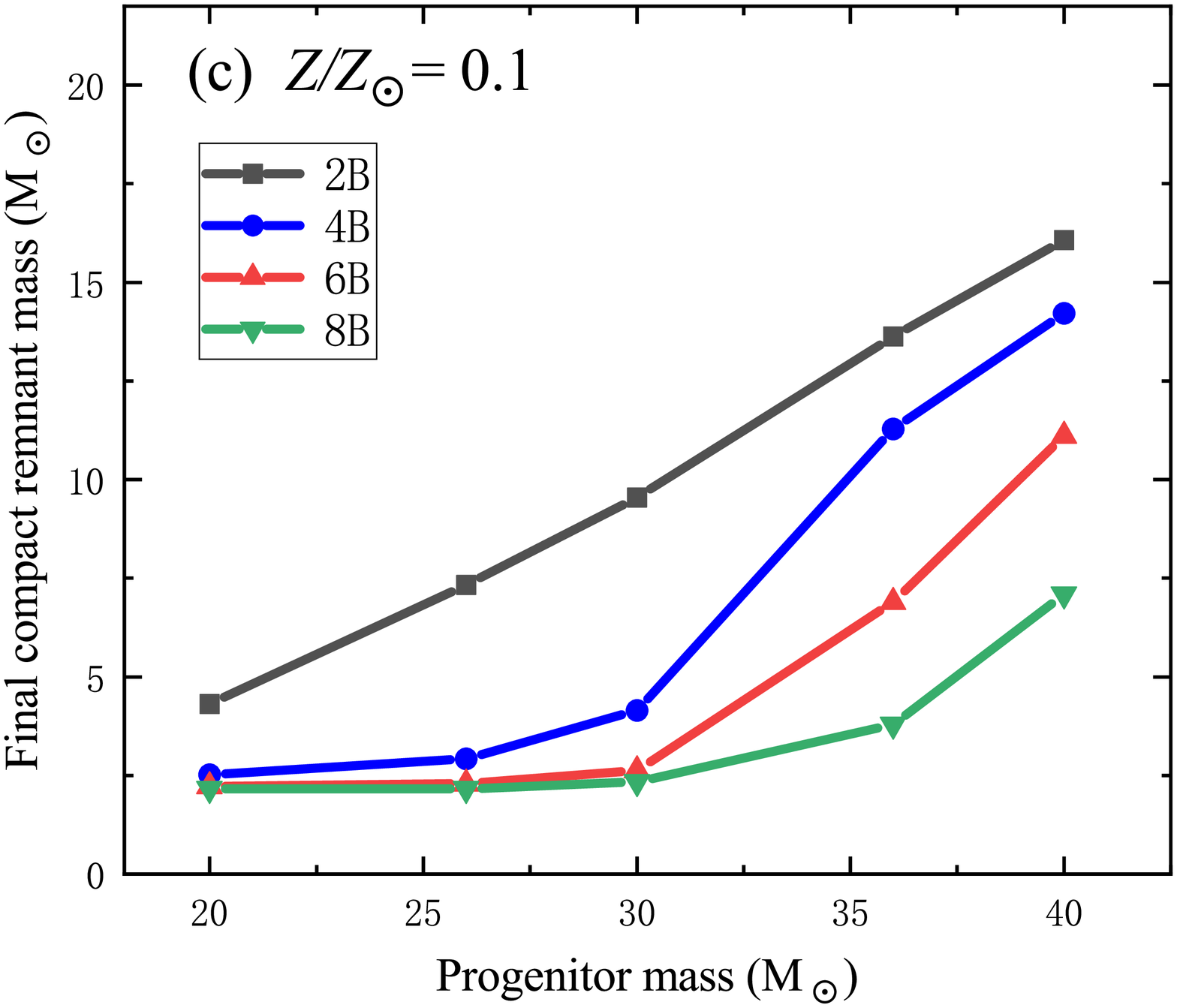}
\includegraphics[width=0.496\linewidth,height=0.43\linewidth]{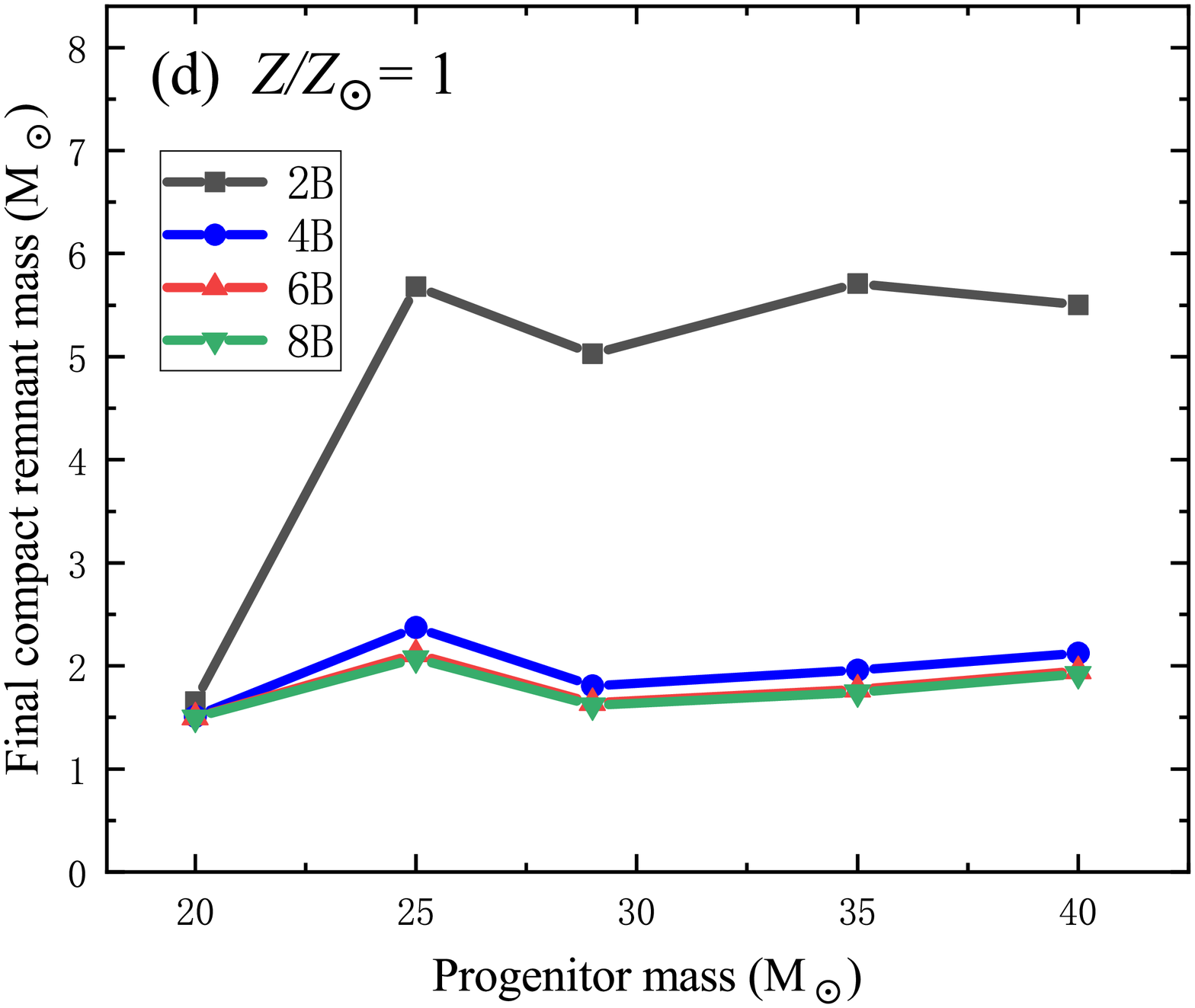}
\caption{Final compact remnant masses in the center of the CCSNe with different explosion energies (2B, 4B, 6B, and 8B, where 1B = $10^{51}$ erg) for the progenitor masses in the range of $20-40~M_\odot$ and with a metallicity of $Z/Z_\odot$ = 0, 0.01, 0.1, and 1.}
\label{fig:mpro-mcr}
\end{figure*}

In the second step, the same outflowing inner boundary condition is set on the same location as the first step ($r=10^9$ cm), but the outer boundary is different, being set on a location far from the star surface ($r=10^{16}$ cm). The medium outside the star is maintained in a constant state with a pressure and density three orders of magnitude lower than the corresponding ones on the star surface. As an essential operation, the additional energy and mass are injected into the innermost cell adjacent to the inner boundary at the beginning of the second step to mimic the outward blast passing through the inner boundary. Here, the injected energy is assumed to have the four values for each case, i.e., 2B, 4B, 6B, and 8B, and the injected mass value is obtained from the recording of inhaled mass during 0.45 s collapse (we assume that there is almost no mass accreted during the outward movement of the piston).

To make the time-step large enough to save the computing time, the grid for the second step has 2000 logarithmic spacing cells, which is much less than the number in the first step. The simulation of all cases has been run until the remnant growth ceased. The maximum duration is about $3\times10^6$ s, which is long enough for the blast to travel over a distance longer than 10 times the star radius (approximately $10^{16}$ cm), a distance at which the gravity of a central compact remnant is relatively weak. The typical time-step amount for simulations of this step is approximately 0.02 s, which is a very high time resolution for the physical process that occurs at a distance far from the central compact remnant.

It should be mentioned that the exploding matter can escape for the outward velocity greater than the escape velocity. On the contrary, the matter with negative velocity should fall into the center. We adopted the criteria by averaging the above bounds as well as that in \citet{Zhang2008}. Furthermore, we also checked the estimates by the time integration of the mass supply rate, which follows $\dot{M} \sim t^{-5/3}$ \citep[e.g.,][]{Chevalier1989,Zhang2008}.

\subsection{Results}

The profiles of the density and velocity at 50, 100, 500, and 1000 s and the time evolutions of the mass supply rate and core mass for $E$ = 2B and 8B with progenitor mass $30~M_\odot$ and $Z/Z_\odot$ = 0.01 are shown in Figure 1. From the comparison of density and velocity profiles between the cases of 2B and 8B, it is obvious that more severe evolution occurs in higher explosions, and is then accompanied by the weaker fallback mass supply and lower core mass growth. If the multiform outflows are inefficient, the mass supply rate can be considered as the net accretion rate, thus we can derive Figure 1(f) from 1(e). It can be noted that the mass supply rate roughly follows $\sim t^{-5/3}$, and whose values in the case of 2B are at least one order of magnitude higher than those in the case of 8B as shown in Figure 1(e). Moreover, by checking the results shown in Figure 1 and Table 1, we insure that our new simulation is correct, as well as those of previous works \citep[e.g.,][]{Zhang2008}, then we can further deduce the reliable conclusions on the lower mass gap.

Figure 2 displays the final compact remnant mass during the fallback processes in the center of the CCSNe with the different explosion energies for the progenitor masses in the range of $20-40~M_\odot$ and metallicity values of $Z/Z_\odot$ = 0, 0.01, 0.1, and 1, corresponding to (a), (b), (c), and (d), respectively. The values of final compact remnant masses are also listed in Table 1.

The trends in Figures 2(a) and (b) are very similar. The core mass of the progenitors with $<30~M_\odot$ and $Z/Z_\odot \lesssim 0.01$ has no significant increase in the cases of the initial explosion energy $\gtrsim$ 4B. For the lower explosion energy $\sim$ 2B, the final remnant mass could increase to tens of the solar mass. If the typically initial explosion energy tends to $\sim$ 2B for all the progenitors, most of the final object mass should be larger than $5~M_\odot$. Furthermore, as shown in Table 1, the residual explosion energy is in the range of $0.01-1.01$B for the cases with the initial explosion energy $\sim$ 2B, and about 6B for $E$ = 8B.

Figure 2(c) displays the final remnant mass for the progenitors with $Z/Z_\odot$ = 0.1. For the massive stars, $>35~M_\odot$, the remnant masses are larger than $5~M_\odot$ for all the progenitors. All the final remnant masses are larger than $5~M_\odot$ for the initial explosion energy of $\sim$ 2B, and only the remnants corresponding to the $40~M_\odot$ progenitors are much lighter than the case of lower metallicities ($Z/Z_\odot<0.01$). The final kinetic energy is in the range of $0.02-0.47$B for the initial explosion energy $\sim$ 2 B, and $2.45-5.25$B for 8B cases as shown in Table 1.

Instead, as shown in Figure 2(d), the core mass of the progenitors with $Z/Z_\odot \sim 1$ cannot grow larger than $\sim 6~M_\odot$ for all values of initial explosion energy. This means the final objects generated in the high-metallicity stars will, unfortunately, fill in the gap. One can imagine that if the solar-metallicity stars are dominant in the nearby universe, we should detect much more compact objects with masses in the range of $2-5~M_\odot$ than the current observations. Thus, according to our model, the existence of the gap indicates that most of the stars in the local universe should be the low-metallicity ($Z/Z_\odot \lesssim 0.1$) ones. Moreover, the metal abundance of the nearby universe can be explained by the nucleosynthesis of the disk outflows from the fallback hyperaccretion in the center of the faint, even failed CCSNe \citep{Liu2020}, instead of the hypernovae.

\section{Applications on the lower mass gap}

\begin{longtable*}{cccccc}
\caption{Data in CCSN simulations}\\
\toprule
\scriptsize{Progenitor mass}  &  \scriptsize{Metallicity} & \scriptsize{Initial explosion energy} & \scriptsize{Initial core mass} & \scriptsize{Residual explosion energy} & \scriptsize{Final compact remnant mass}   \\
($M_\odot$)  & ($Z_\odot$)  & (B)   & ($M_\odot$)    &  (B)   & ($M_\odot$) \\
\midrule

40	&	0	&	2	&	2.08 	&	0.03 	&	35.01 	\\
40	&	0	&	4	&	2.08 	&	0.96 	&	15.89 	\\
40	&	0	&	6	&	2.08 	&	2.82 	&	9.27 	\\
40	&	0	&	8	&	2.08 	&	5.41 	&	5.02 	\\
35	&	0	&	2	&	1.94 	&	0.19 	&	19.26 	\\
35	&	0	&	4	&	1.94 	&	1.67 	&	9.90 	\\
35	&	0	&	6	&	1.94 	&	4.25 	&	3.94 	\\
35	&	0	&	8	&	1.94 	&	6.47 	&	2.70 	\\
30	&	0	&	2	&	1.93 	&	0.78 	&	9.97 	\\
30	&	0	&	4	&	1.93 	&	2.62 	&	2.41 	\\
30	&	0	&	6	&	1.93 	&	4.52 	&	2.09 	\\
30	&	0	&	8	&	1.93 	&	6.31 	&	2.02 	\\
25	&	0	&	2	&	1.88 	&	1.09 	&	2.92 	\\
25	&	0	&	4	&	1.88 	&	3.02 	&	1.99 	\\
25	&	0	&	6	&	1.88 	&	4.87 	&	1.94 	\\
25	&	0	&	8	&	1.88 	&	6.57 	&	1.93 	\\
20	&	0	&	2	&	1.92 	&	1.01 	&	3.06 	\\
20	&	0	&	4	&	1.92 	&	2.54 	&	2.49 	\\
20	&	0	&	6	&	1.92 	&	4.47 	&	2.01 	\\
20	&	0	&	8	&	1.92 	&	6.22 	&	1.99 	\\
40	&	0.01	&	2	&	1.96 	&	0.01 	&	40.00 	\\
40	&	0.01	&	4	&	1.96 	&	0.53 	&	16.30 	\\
40	&	0.01	&	6	&	1.96 	&	1.90 	&	11.86 	\\
40	&	0.01	&	8	&	1.96 	&	4.11 	&	7.03 	\\
36	&	0.01	&	2	&	2.28 	&	0.06 	&	24.00 	\\
36	&	0.01	&	4	&	2.28 	&	0.98 	&	12.48 	\\
36	&	0.01	&	6	&	2.28 	&	3.25 	&	5.86 	\\
36	&	0.01	&	8	&	2.28 	&	5.62 	&	3.55 	\\
30	&	0.01	&	2	&	2.09 	&	0.44 	&	11.15 	\\
30	&	0.01	&	4	&	2.09 	&	2.62 	&	3.52 	\\
30	&	0.01	&	6	&	2.09 	&	4.85 	&	2.48 	\\
30	&	0.01	&	8	&	2.09 	&	6.85 	&	2.27 	\\
26	&	0.01	&	2	&	1.91 	&	0.52 	&	8.52 	\\
26	&	0.01	&	4	&	1.91 	&	2.73 	&	2.70 	\\
26	&	0.01	&	6	&	1.91 	&	4.76 	&	2.18 	\\
26	&	0.01	&	8	&	1.91 	&	6.68 	&	2.06 	\\
20	&	0.01	&	2	&	2.05 	&	0.67 	&	4.19 	\\
20	&	0.01	&	4	&	2.05 	&	2.47 	&	2.38 	\\
20	&	0.01	&	6	&	2.05 	&	4.32 	&	2.19 	\\
20	&	0.01	&	8	&	2.05 	&	6.31 	&	2.15 	\\
40	&	0.1	&	2	&	1.90 	&	0.02 	&	16.07 	\\
40	&	0.1	&	4	&	1.90 	&	0.33 	&	14.21 	\\
40	&	0.1	&	6	&	1.90 	&	1.45 	&	11.11 	\\
40	&	0.1	&	8	&	1.90 	&	2.45 	&	7.10 	\\
36	&	0.1	&	2	&	2.18 	&	0.04 	&	13.63 	\\
36	&	0.1	&	4	&	2.18 	&	0.42 	&	11.28 	\\
36	&	0.1	&	6	&	2.18 	&	2.13 	&	6.89 	\\
36	&	0.1	&	8	&	2.18 	&	3.41 	&	3.79 	\\
30	&	0.1	&	2	&	2.13 	&	0.15 	&	9.55 	\\
30	&	0.1	&	4	&	2.13 	&	1.55 	&	4.15 	\\
30	&	0.1	&	6	&	2.13 	&	3.62 	&	2.63 	\\
30	&	0.1	&	8	&	2.13 	&	5.25 	&	2.33 	\\
26	&	0.1	&	2	&	2.02 	&	0.20 	&	7.33 	\\
26	&	0.1	&	4	&	2.02 	&	1.85 	&	2.92 	\\
26	&	0.1	&	6	&	2.02 	&	2.60 	&	2.29 	\\
26	&	0.1	&	8	&	2.02 	&	4.90 	&	2.16 	\\
20	&	0.1	&	2	&	2.06 	&	0.47 	&	4.31 	\\
20	&	0.1	&	4	&	2.06 	&	1.94 	&	2.52 	\\
20	&	0.1	&	6	&	2.06 	&	2.22 	&	2.22 	\\
20	&	0.1	&	8	&	2.06 	&	4.74 	&	2.17 	\\
40	&	1	&	2	&	1.84 	&	0.87	&	5.50 	\\
40	&	1	&	4	&	1.84 	&	3.43	&	2.12 	\\
40	&	1	&	6	&	1.84 	&	5.45	&	1.95 	\\
40	&	1	&	8	&	1.84 	&	7.32	&	1.92 	\\
35	&	1	&	2	&	1.67 	&	0.95	&	5.71 	\\
35	&	1	&	4	&	1.67 	&	3.19	&	1.96 	\\
35	&	1	&	6	&	1.67 	&	5.16	&	1.77 	\\
35	&	1	&	8	&	1.67 	&	7.05	&	1.74 	\\
29	&	1	&	2	&	1.54 	&	1.06	&	5.03 	\\
29	&	1	&	4	&	1.54 	&	3.22	&	1.81 	\\
29	&	1	&	6	&	1.54 	&	5.01	&	1.64 	\\
29	&	1	&	8	&	1.54 	&	6.75	&	1.61 	\\
25	&	1	&	2	&	1.97 	&	1.12	&	5.68 	\\
25	&	1	&	4	&	1.97 	&	2.32	&	2.37 	\\
25	&	1	&	6	&	1.97 	&	4.02	&	2.12 	\\
25	&	1	&	8	&	1.97 	&	5.51	&	2.07 	\\
20	&	1	&	2	&	1.46 	&	1.23	&	1.65 	\\
20	&	1	&	4	&	1.46 	&	2.86	&	1.51 	\\
20	&	1	&	6	&	1.46 	&	4.38	&	1.50 	\\
20	&	1	&	8	&	1.46 	&	5.81	&	1.50 	\\

\bottomrule																																																																		\end{longtable*}

It is generally considered that metallicity is one of the crucial factors in star formation. Most of the massive Population III and II stars should be the metal-poor stars \citep[e.g.,][]{Bromm2013}. Recently, \citet{Cordoni2020} reported tens of extremely metal-poor stars in the Galactic plane from Gaia DR2, which imply that the lower-metallicity stars might be far more than expected. Considering the above results from Figure 2, we use the case of $Z/Z_\odot$ =0.01 to estimate the distribution of the compact objects produced by $20-40~M_\odot$ stars.

Based on the simulation results in Figure 2(b), we derive the fitting formula of the relation on the compact remnant mass, progenitor mass, and initial explosion energy as follows:
\begin{equation}
\log m_{\rm cr} = \left\{
\begin{array}{rcl}
-1.13 + 1.64 &\log& m_{\rm pro} - 1.01 \log e, \\&&20 \leq m_{\rm pro} < 30 \\
-5.53 + 4.68 &\log& m_{\rm pro} - 1.24 \log e, \\&&30 \leq m_{\rm pro} \leq 40 \\
\end{array} \right.
\end{equation}
where $m_{\rm cr}=M_{\rm cr}/M_\odot$, $m_{\rm pro}=M_{\rm pro}/M_\odot$, and $e=E/1\rm B$ are the dimensionless remnant mass, progenitor mass, and initial explosion energy, respectively.

Here, we adopt the initial mass function (IMF) of \cite{Salpeter1955} to generate mass distributions of progenitor stars. The progenitor metallicity effects are reasonably ignored. Thus, by using the above fitting formula, we can connect the distributions of progenitors to those of the final compact remnants. We assign the initial explosion energy of each progenitor according to a random Gaussian distribution with a standard deviation of 1B. For the mean of the Gaussian distribution, we consider three cases, i.e., mean explosion energy $\overline{E}$ = 2B, 4B, and 8B. Motivated by the limit of the gravitational energy of the cores in collapsars, the Gaussian distribution is truncated to the range of $0.1-10$B (i.e., we consider that the luminously observable CCSNe are totally scarce). The resulting remnant mass is determined by the mass of the progenitor star and the initial explosion energy. The probability density distributions of the compact remnant mass for three $\overline{E}$ are shown in Figure 3. Of course, the different progenitors might correspond to different mean explosion energies. Since there is a serious limitation in the typical energy for the different progenitors, we have to assume that their values might be close to the progenitor masses in the range of $20-40~M_\odot$. Moreover, as shown in Figure 2(b), the final remnant mass of the massive progenitors ($\gtrsim 35~M_\odot$) is larger than $5~M_\odot$, which does not influence the gap.

From Figure 2(b), it is easy to notice that the initial cores with the progenitors $\lesssim 35~M_\odot$ cannot efficiently grow from $< 2~M_\odot$ to $>5~M_\odot$ to stride the gap in the more violent explosions, so most of the final remnants remain on the left-hand side of the gap in Figure 3. Furthermore, the cores grow in the range of the gap when the initial explosion energy approaches 4B. For the lower explosion energy, most of the core masses are increased beyond $5~M_\odot$. It seems that regardless of whether the higher or lower explosion energy is typical, the gap might exist. If drastic explosions are universal, the number of detectable CCSNe should be far more than the current observations, because the final kinetic energy is larger than 4B in the cases of $E$ = 8B and $Z/Z_\odot$ = 0.01 as shown in Table 1. Thus, the typical explosions, $\lesssim$ 2B, are entirely reasonable. The progenitors with masses $\lesssim 20~M_\odot$ should produce a mass of NSs, and $>5~M_\odot$ BHs would be generated by the stars with $>40~M_\odot$ but with the additional contribution of the progenitors with $20-40~M_\odot$. Therefore, we suggest that the lower mass gap probably exists and that its shape depends on the typical explosion energies for all sorts of progenitor stars. The final remnants that originated from the high-metallicity stars might fill the gap a little. Furthermore, in the binary system, the companion might influence the evolution of the stars during the interactions, even mergers \citep[e.g.,][]{Fryer2012}.

\begin{figure}[t]
\centering
\includegraphics[width=1.0\linewidth]{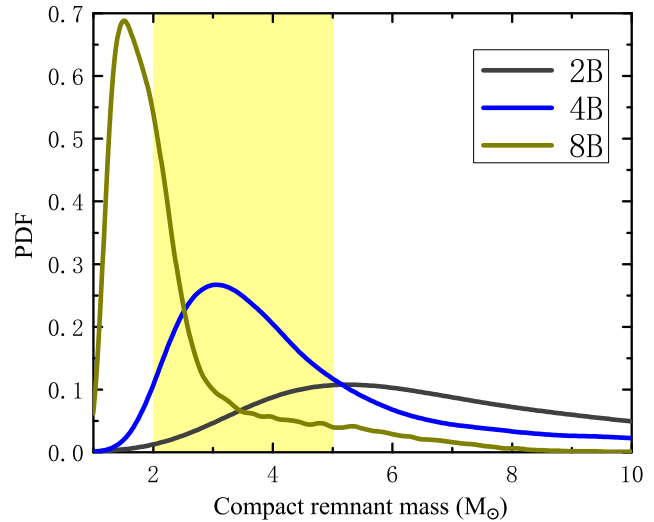}
\caption{Probability density distributions of the compact remnant mass for three typical CCSN initial explosion energies.}
\label{fig:distribution}
\end{figure}

\section{Conclusions and discussion}

In this paper, we propose the possibility of the formation of the lower mass gap by considering the fallback mechanism in the CCSN scenario. If the typical initial explosion energy of the CCSNe is much lower than that of the observable ones, the cores of the progenitors with a mass of $<20~M_\odot$ cannot collapse into BHs, and those in the range of $20-40~M_\odot$ collapse into BHs with a mass of $>5~M_\odot$ due to the fallback accretion; thus, the lower mass gap will naturally exist. Only some of the solar-metallicity stars still generate compact objects in the mass gap. In addition, low-metallicity massive progenitors ($> 40~M_\odot$) with a low explosion energy could contribute to the distributions of the BHs with tens of solar mass.

In the above framework, the shape of the gap is mainly determined by the typical explosion energy of the CCSNe from 20 to 40 $M_\odot$. However, the physics of CCSNe still remains uncertain in terms of the explosion mechanism, the morphology, the effects of the progenitors, and so on. Once the further multimessenger observations and more detailed simulations on CCSNe adequately supply the information on the mass gap, the properties of the CCSNe, including their progenitors, could be well constrained. In addition, there is another important CCSN branch named ``failed'' SNe \citep[e.g.,][]{MacFadyen1999}. \citet{Kobayashi2020} estimated that their proportion is about 50$\%$ early in the history of the Galaxy, which might be a potential supplement for our results.

In our previous works \citep[e.g.,][]{Liu2015,Liu2018,Song2016,Song2019}, by using the observations on the long-duration gamma-ray bursts (LGRBs), we investigated the feasibility of the neutrino annihilation process \citep[e.g.,][]{Liu2017} and Blandford-Znajek jets \citep{Blandford1977} launched by the BH hyperaccretion systems in the collapsar scenario. Since the BH mass and spin undergo violent evolution during the hyperaccretion process, we will further test the BH growth by the LGRB observations to study the existence of the lower mass gap.

\acknowledgments
We thank Prof. Ang Li, Prof. Yu Gao, and Dr. Cui-Ying Song for helpful discussions. This work was supported by the National Natural Science Foundation of China under grants 11822304 and 11973002 and the Natural Science Foundation of Fujian Province of China under grant 2018J01007.

\software{Athena++}

\end{document}